
\rightline{IASSNS-HEP-94/26}
\smallskip
\rightline{April, 1994}
\leftline{\bf Composite leptons and quarks constructed as triply occupied}
\leftline{{\bf quasiparticles in quaternionic quantum
mechanics}\footnote{$^*$}{Submitted for publication to Physics Letters
B}}
\bigskip
\leftline{Stephen L. Adler}
\leftline{\it Institute for Advanced Study, Olden Lane, Princeton NJ
08540 USA}
\bigskip
\noindent We propose a set of rules for constructing composite
leptons and quarks
as triply occupied quasiparticles, in the quaternionic quantum mechanics
of a pair of Harari-Shupe preons $T$ and $V$.  The composites fall into
two classes, those with totally antisymmetric internal wave functions,
and those with internal wave functions of mixed symmetry.  The mixed
symmetry states consist of precisely the three spin 1/2 quark lepton
families used in the standard model (48 particle states, {\it
not}
counting the doubling arising from antiparticles),
plus one doublet of spin 3/2 quarks (24 particle states).
The antisymmetric states consist of a set of spin 3/2 leptonic states
with charges as in a standard model family (16 particle states), and a spin 1/2
leptonic fractionally charged doublet (4 particle states).  We sketch ideas for
deriving our rules from a fundamental quaternionic preonic field theory.
\bigskip
\bigskip
Although the repetitive family structure of the standard model
leptons and quarks is strongly suggestive of further substructure, it
has proved difficult to formulate economical preon model candidates.
A useful step in this direction was taken in 1979 by Harari [1] and
Shupe [2], who proposed a set of rules for generating the states
in a single lepton-quark family (including the color tripling of the
quarks)
as triples constructed from two fundamental preons $T$ and $V$, with
charges 1/3 and 0 respectively.  In order to generate the color states
this way, the Harari-Shupe rules require that $TTV$, $TVT$, and $VTT$ be
counted as distinct states, which is not possible within standard
quantum mechanics, but which I suggested [3] might be possible within
quaternionic quantum mechanics, where the multiplication of Hilbert
space scalars is noncommutative.  I later noted [4] that if one includes
spin states in the enumeration, there is the possibility of generating
family structure as well as color structure dynamically; however, in the
naive formulation of Ref. [4] (which proceeds by permutation
of character strings, as in Refs. [1,2]) one gets only two spin 1/2
families together with a spin 3/2 family, and there seemed to be no
plausible mechanism to generate the additional spin 1/2 states needed
for a third family.

Despite the lack of progress in further developing the Harari-Shupe
proposal, in the intervening years there has been considerable progress
in the analysis of the structure of quaternionic quantum mechanics
(see, e.g. [5,6]), which is summarized and considerably expanded on in my book
[7].
In this Letter, I use the  methods developed in Ref. [7] for the second
quantization of many-body systems in quaternionic quantum mechanics, and
in particular the construction given there of quasiparticles with
quaternionic wave functions, to take a new look at the issue of preon
models.  We shall see that when the structure of quaternionic
quasiparticle wave
functions is taken into account by a plausible set of rules, the
proposal of Refs. [1-4] to generate color {\it and} family structure from
two fundamental preons can in fact be realized.  We begin by stating
the rules and showing  how they are used to enumerate three quasiparticle
composite states; we then turn to a discussion of how the rules might
be derived from a fundamental relativistic preonic theory.

Let $p_n(\vec r \,)$ and $p^{\dagger}_n (\vec r \,)$
be ordinary fermionic annihilation
and creation operators  for the preonic states,
$$ \{ p_n \}=\{ t_{\uparrow},t_{\downarrow},v_{\uparrow}, \break
 v_{\downarrow} \}, \eqno(1a) $$
$$\eqalign{
\{ p_m(\vec r \,),p_n(\vec r \,' ) \} =&0, \cr
\{ p^{\dagger}_m(\vec r \,),p^{\dagger}_n(\vec r\,' ) \} =&0, \cr
\{ p_m(\vec r \,), p^{\dagger}_n(\vec r\,') \} =&\delta_{mn} \break
\delta^3(\vec r - \vec r\,') , \cr
}
\eqno(1b)$$
where $t$ and $v$ carry charges 1/3 and 0 respectively and the arrows
indicate the spin.  It is shown in Ref. [7] that these creation and
annihilation operators can be treated as real quantities with respect
to an appropriately defined quaternion algebra (a left-acting or
operator quaternion algebra) $1,E_1,E_2,E_3$, which obeys
$$ E_A E_B = -\delta_{AB} + \sum_C \epsilon_{ABC} E_C, \eqno(2a)$$
and which can be reexpressed in an obvious vector notation (with real
number vectors $\vec a, \vec b $ ) as
$$\vec E \cdot \vec a \> \vec E \cdot \vec b = -\vec a \cdot \break
 \vec b + \vec a \times \vec b \cdot \vec E. \eqno(2b) $$
According to Ref. [7], and as discussed below, quasiparticle
annihilation operators are formed as superpositions of ordinary
annihilation operators, with quaternion-valued wave functions as
coefficients.  As our first rule, we assume
\smallskip
\noindent {\it {\bf Rule 1.}  The wave function appearing in the quasiparticle
operators
is assumed to be quaternion imaginary, and to have nonvanishing and
linearly independent components along the three quaternion units $\vec
E$.  The preon binding forces, and thus the ground state wave function,
are assumed also to be flavor (i.e., $t,v$) and spin (i.e.,
$\uparrow,\downarrow$) independent.}
\smallskip
Hence the annihilation operator $P_n(\vec R \,)$
for a quasiparticle  located at $\vec R$ has the form
$$P_n(\vec R \,)=\int d^3r \> \vec a(\vec r \, ) \cdot \vec E \> \break
p_n(\vec r+ \vec R \,), \eqno(3a)$$
which can be written in abbreviated form as
$$P_n=\sum_1 \vec a(1) \cdot \vec E \> p_n(1). \eqno(3b) $$
In Table 1 we explicitly write out the quasiparticle operators for the
two spin states of the two preonic flavor states, in both the full
notation of Eq. (3a) and the abbreviated notation of Eq. (3b).
We shall assume that the wave function $\vec a(\vec r \,)$ has support only for
$|\vec r \,|$ of order the preonic length scale, which we assume to be much
smaller than length scales characterizing standard model physics, and
that the wave function is unit normalized,
$$ \int d^3r \> |\vec a(\vec r\, )|^2 =1. \eqno (3c) $$

The quasiparticle operator defined by Eqs. (3a,b) has a number of
interesting properties.  First of all, a simple calculation shows that
any quasiparticle state can at most be triply occupied, since the
fourth power of a quasiparticle operator
vanishes [7] (this property, first noted in a different
context by Govorkov [8], holds even when the internal wave function
$\vec a$ has a real part):
$$P_n(\vec R \,)^4=0. \eqno(4a)$$
Also, the quasiparticles satisfy parastatistics-like commutation relations
[8]
(here the assumption of a quaternion imaginary internal wave function is
needed):
$$\eqalign{
[P_{\ell},[P_m,P_n]]=&0, \> {\rm any} \> \ell,m,n , \cr
[P^{\dagger}_{\ell},[P_m,P^{\dagger}_n]]=&-2  \delta_{\ell m} \break
\sum_{1,2} \vec a(1) \cdot \vec E \> \vec a(1) \cdot \vec a(2) \break
p^{\dagger}_n(2) .\cr
}
\eqno(4b)  $$
[If one were to replace Eq. (3c) by the stronger condition
$$\int d^3r \> a_A(\vec r \,)a_B(\vec r \,)=\delta_{AB}/3, \eqno(4c)$$
then the second line of Eq. (4b) would become the parastatistics
commutator
$$[P^{\dagger}_{\ell},[P_m,P^{\dagger}_n]]=(2/3)  \delta_{\ell m} \break
P^{\dagger}_n . \eqno(4d)$$
However, we do not assume Eqs. (4c,d) in what follows.]
Eqs.(4a,b) suggest that states of three quasiparticles will play a
special role, and we identify them  as composite leptons and quarks,
which are formed and classified according to the following further
rules:
\smallskip
\noindent {\it {\bf Rule 2.}  Composite fermion states are identified with the
{\bf quaternion real
components}  of the independent products of three quasiparticle operators
drawn from Table 1. In other words, a product $C=P_{\ell}P_mP_n$ which is
already
quaternion real is identified as a composite fermion operator, while
if $C$ has the quaternion imaginary form $C=\vec C \cdot \vec E$, then
the real components $C_1, C_2, C_3$ are each identified as an
independent composite fermion operator.}
\smallskip
\noindent {\it {\bf Rule 3.}
When the number of composite states is tripled as a result of the inequivalence
of different orderings of $t$ and $v$, the composites are identified as
colored quark states, corresponding to the fact that the underlying
preonic forces can cause transitions between $t$ and $v$.  When the number of
composite states is tripled as a result of the inequivalence of
different orderings of the spin labels $\uparrow$ and $\downarrow$, the
composites are identified as states in different
families, corresponding to the fact that the underlying preonic forces
are assumed spin independent, and so do not cause transitions between
$\uparrow$ and $\downarrow$.}
\smallskip
The enumeration of the independent products of three
quasiparticle operators proceeds much as the enumeration of states in
the quark model [9,10], and is expedited by the use of a number of
simple identities which follow from the defining equations given above.
First of all, from Eq. (2b) we immediately get
$$P_{\ell} P_m=\sum_{1,2} [-\vec a(1) \cdot \vec a(2) \break
+ \vec a(1) \times \vec a(2) \cdot \vec E ] \> \break
 p_{\ell}(1) p_m(2) , \eqno(5a) $$
allowing us to easily evaluate the successive products of quasiparticle
operators.  The classification of products according to their
reality properties under
quaternion conjugation (indicated by a bar $\bar{} \>$) is facilitated by
noting that for any $\ell, m, n$ we have
$$ \overline  {P_{\ell} P_m P_n}= P_n P_m P_{\ell}, \eqno(5b) $$
which follows from the facts that (i) the quaternion conjugate of a product
of factors is the product of the conjugates of the factors in reverse
order, (ii) the conjugate of an imaginary quaternion is minus itself,
and (iii) reverse ordering the product of
three fermionic annihilation operators just reverses the sign of the
product.  Finally, setting $n=\ell$ in the first line of Eq. (4b)
gives the repeatedly used
identity, again valid for any $\ell, m$,
$$P_{\ell} P_{\ell}  P_m + P_m P_{\ell} P_{\ell} = \break
2 P_{\ell} P_m P_{\ell}.  \eqno(5c) $$
The most complicated enumeration is for the charge 2/3, spin
and helicity 1/2 states, which can have the three charge structures
$TTV+VTT$, $TVT$, and $TTV-VTT$, combined with the two
spin structures $\uparrow \uparrow \downarrow + \downarrow \uparrow \uparrow
-2 \uparrow \downarrow \uparrow$ and $\uparrow \uparrow \downarrow
- \downarrow \uparrow \uparrow$, giving six possibilities in all.  One
readily finds, using Eq. (5b), that $TTV+VTT$ and $TVT$ combined with
$\uparrow \uparrow \downarrow-\downarrow \uparrow \uparrow$ are
imaginary, as is $TTV-VTT$ combined with $\uparrow \uparrow \downarrow
+\downarrow \uparrow \uparrow - 2 \uparrow \downarrow \uparrow$, and
these correspond to the 3 families of charge 2/3 quarks shown in Table 3.
The remaining three combinations are real, but
only one is linearly independent after using Eqs. (4b) and (5c), and
corresponds to the spin 1/2 lepton in Table 2.   For the charge 1,
spin and helicity 1/2 states, there is only one charge structure $TTT$,
which when combined with the spin structure $\uparrow \uparrow \downarrow
- \downarrow \uparrow \uparrow$ gives one imaginary composite,
corresponding to the 3 families of charge 1 leptons in Table 3; the
contribution of the spin structure $\uparrow \uparrow \downarrow
+ \downarrow \uparrow \uparrow
-2 \uparrow \downarrow \uparrow$ vanishes in this case by Eq. (5c).  The
computations in the spin 3/2 cases, which involve totally symmetric spin
structures, are similar.

The results of the calculation are summarized in Tables 2 and 3.
Table 2 lists all states with a totally antisymmetric internal
wave function proportional to $\vec a(1) \times \vec a(2) \cdot \vec
a(3)$.  Table 3 lists all states with a mixed symmetry internal
wave function proportional to $\vec a(1) \cdot \vec a(2) a_A(3)$
or to the two structures obtained from this one by permuting the labels
1,2,3 for fixed $A$, with $A$ an index which also takes the values 1,2,3, each
value of $A$ being counted (by Rule 2) as a distinct state.
Although there is no dynamics for mass generation in the model,
experience with the quark model suggests that  states with similar
internal wave
functions should have roughly similar masses, at least when viewed on
a preonic energy scale, and that the states with mixed symmetry internal
wave functions should be lighter than those which are totally
antisymmetric.  Thus, it is encouraging that the spin 1/2 content of the
mixed symmetry states corresponds, when interpreted by Rule 3, with
precisely the content of the fermions used in the standard model.
One striking feature of the spin 1/2 wave functions in Table 3 is that
all three leptons, and the first two sets of quarks, have spin
structure $\uparrow \uparrow \downarrow-\downarrow \uparrow \uparrow$,
while the third set of quarks has spin wave function $\uparrow \uparrow
\downarrow + \downarrow \uparrow \uparrow - 2 \uparrow \downarrow
\uparrow$.  Hence a strong mass operator dependence on spin state would
result in a large mass splitting between the third set of quarks and the
remaining quarks and leptons, roughly corresponding to what is
experimentally observed for the quarks of the third family
(particularly if the measure of the zeroth order quark mass of a family
is taken as a geometric or arithmetic mean of the masses of the two
charge states in that family.)

If the spin 3/2 mixed symmetry quarks are nearby, they may be observable
at LEP and the Tevatron, or at the LHC.  In all likelihood, they should
be unstable against magnetic dipole electromagnetic decay into spin 1/2
quarks, and hence should not be seen directly (through new types of
mesons),  but rather should only appear indirectly through enhanced
production of standard model quarks (and mesons) in certain channels.
If preliminary
hints of an excess branching ratio into $b$ quarks at LEP [11] survive the
accumulation of better statistics, then production of the spin 3/2
quarks of Table 3 could be considered as a possible explanation.
However, since the spin 3/2 quarks cannot cancel chiral anomalies among
themselves, they must have vector rather than chiral electroweak
currents.  This permits them to have mass terms even in the absence of
electroweak spontaneous symmetry breaking, and so they could in
principle have masses much larger than those of the standard model spin
1/2 fermions, which have chiral electroweak currents and must get
their masses through the electroweak Higgs mechanism.
Whether the states of Table 2 will be visible in the near future is hard
to assess without a detailed dynamics; they could in principle
lie at the preonic mass scale, which in turn may well be as high as the
GUT scale, in which case they may never be directly
observable in accelerator experiments.

We note that when any two of the
composite annihilation operators of Tables 2 and 3
are anticommuted, one gets zero as expected for canonical fermions.
However, when a composite annihilation operator is anticommuted with
the adjoint of another, one gets both a c-number term and operator terms.
The operator terms contribute significantly to vacuum
expectations of products of
composite operators only when there is an appreciable probability for
composite particles to approach within the preonic distance scale, in which
case
their substructure becomes visible; at energies far below the preonic
scale the operator terms in the anticommutators can be neglected.
For general composites $C_{\ell}, C_m$ from Tables 2 and 3, the c-number
parts of the anticommutators have the form
$$\{ C_{\ell}(\vec R \>),C^{\dagger}_m(\vec R' \>) \}= \break
\delta^3(\vec R - \vec R' \>) c_{\ell m} , \eqno(6)$$
with $c_{\ell m}$ a real, symmetric matrix which is not in general in diagonal
form.  To construct the independent degrees of freedom, one must form
linear combinations of the composites within each spin and charge sector
 of the Tables using the real,
orthogonal matrix which diagonalizes the matrix $c_{\ell m}$ for that
sector, and then do an appropriate renormalization to get a unit
anticommutator.  The considerations of this paragraph
all have analogs in the ordinary quark
model.

Let us now turn to a brief discussion of how one could try to justify the
rules given above from a more fundamental preonic theory.  In a recent
paper [12] I proposed a generalized quantum dynamics which can apply to
quaternionic as well as standard complex quantum field theories, and
which leads to equations of motion that are invariant under operator
valued gauge transformations.  I noted there (see also Chapts. 12 and 13
of Ref.
[7]) that the minimal fermionic model with maximal (two-sided) operator
gauge invariance necessarily has {\it two} fermion fields, basically
because in order to get the right hermiticity properties without breaking one
of the gauge groups, one must use
the two dimensional real matrix representation  of the
imaginary unit (given in standard Pauli matrix form as -$i \sigma_2$) in
constructing the fermionic Lagrangian.  Thus the minimal fermionic
 model contains the
two fields $t$ and $v$ which are needed to make lepton quark composites.
In addition, the model of Ref. [12] contains only vector couplings to its
gauge gluons, and has  an (anti)symmetrical structure in the two
fundamental fermions.  The
vector-like structure means that the forces binding the preons
are spin-independent, and the fermionic symmetry can plausible lead to binding
forces which are flavor-symmetric, as assumed in Rule 1.  Finally, the
model of Ref. [12] has a global chiral invariance (although it cannot be
decoupled into noninteracting chiral components of the fields), and so
it is possible for the lowest lying composites formed from the preons to
have
zero mass on the preonic mass scale, as is needed in a physically realistic
preon model.  For these reasons, it is natural to conjecture [7] that
the minimal fermionic model of Ref. [12] gives the underlying
relativistic preon dynamics.

Although there are many open questions in the generalized dynamics
discussed in Ref. [12], let us suppose that there is a regime in which
this dynamics is represented by a unitary operator dynamics in Hilbert
space.  We can than invoke a general result proved in Ref. [7] (see Ref.
[13] for an earlier version), which asserts that for a general quaternionic
operator Hamiltonian dynamics, except in the strictly
zero energy sector the $S$-matrix in quaternionic Hilbert space is
complex. This means that the asymptotic state dynamics will be
represented by an {\it effective} complex quantum field theory acting on
the asymptotic particle states. To justify Rule 2, one would then have
to show that the asymptotic particles correspond to three quasiparticle
composites, with an effective anti-self-adjoint time development
operator of the form
$$\tilde H= tr_E[\partial C^{\dagger}/{\partial t} \>C - \break
C^{\dagger} \> \partial C/ {\partial t}], \eqno(7) $$
with $tr_E$ denoting a trace over the operator quaternion algebra
spanned by $\vec E$.  Given the role of traces in the dynamics
formulated in Ref. [12], this form for the asymptotic dynamics is
plausible.

Let us next consider the binding of preons into composites.  Since the
model of Ref. [12] has a QCD-like gauge structure [based on a
quaternionic extension of an $SU(2) \times SU(2)$ gauge theory], it is
not unreasonable to suppose that the formation of composite bound states
can be treated much as in QCD.  In QCD, although the light mesons are
actually
highly relativistic bound states, one finds that the classification of
the low-lying hadronic states can be successfully carried out in the  non-
relativistic quark model [9], in which quark binding is treated in the shell
model approximation.  In the shell model, which is based on the Hartree
or self-consistent field approximation,
one assumes that
each particle moves independently in a potential centered on the center
of mass of the overall system.   Taken over to our quaternionic preon
model, the shell model anti-self-adjoint Hamiltonian $\tilde H_3$ for
the binding of three preons, with
spin and flavor independent forces and with the center of mass chosen as
the origin, takes the form
$$
\eqalign{
\tilde H_3=&\sum_{n=1}^3 \tilde H(\vec r_n), \cr
\tilde H(\vec r \,)=&\{ (-I/2M)[\vec \nabla_{\vec r} - I \break
\vec A(\vec r \,)]^2 + \tilde U(\vec r \,) \}, \cr
}
\eqno(8a) $$
with M the preon effective mass and with $\vec A$  and $\tilde U$
respectively a real vector potential and a quaternion imaginary scalar
potential which are both functions of the preon coordinate $\vec r$.
Since Eq. (8a) is the sum of identical one body Hamiltonians for the
three particles, it can be rewritten in Fock space as [7]
$$\tilde H_F=\int d^3r \> p^{\dagger}(\vec r \,) \tilde H(\vec r \,) \break
p(\vec r \,), \eqno(8b) $$
with $\tilde H(\vec r \,)$ the one body Hamiltonian defined in Eq. (8a).
Now let $a_{\kappa}(\vec r \,)$ be a complete orthonormal
set of one particle energy
eigenstates of the one body Hamiltonian, obeying
$$\tilde H(\vec r \,) a_{\kappa}(\vec r \,) =a_{\kappa}(\vec r \,) \break
I E_{\kappa}. \eqno(8c) $$
Then defining the {\it quasiparticle operator} $p_{\kappa}$ and its
adjoint by
$$
p_{\kappa}=\int d^3r \>\overline{a_{\kappa}} (\vec r \,) p(\vec r \,) \break
,\>\>\>\break
p^{\dagger}_{\kappa}=\int d^3r \> p^{\dagger}(\vec r \,) \break
a_{\kappa} (\vec r \,), \eqno(9a) $$
some simple algebra [7] (which parallels the corresponding derivation in
 standard complex quantum
mechanics ) shows that $\tilde H_F$ can be rewritten as
$$\tilde H_F=\sum_{\kappa} p^{\dagger}_{\kappa} I E_{\kappa} \break
p_{\kappa}.  \eqno(9b)$$
As shown in Ref. [7], the energy eigenstates in the one particle sector
are exactly created by the quasiparticle operators
$p^{\dagger}_{\kappa}\>$, but since the latter do
not obey canonical commutators
because of the noncommutativity of quaternionic wave functions, they
do not behave as creation operators for
independent quasiparticles in sectors with more
than one particle.

Nevertheless, let us assume that it is reasonable to
approximate the creation
operator for the ground
state in the three particle sector as a product of three ground state
quasiparticle
creation operators.  We then get the product recipe for creating
composites given in Rule 2. In general the ground state wave function
$a_{0}(\vec r \,)$ is not quaternion imaginary, but we now observe that
if we rewrite it in symplectic component form,
$$a_{0}(\vec r \,)=a_{0}(\vec r \,)_{\alpha} + J a_{0}(\vec r \,)\break
_{\beta}, \eqno(10a) $$
with the $\alpha,\beta$ components in the complex subalgebra spanned by
1 and $I$, we can always find a complex phase $\zeta(\vec r \,)$ which
makes $\zeta(\vec r \,) a_{0}(\vec r \,)$ quaternion imaginary.  (Simply take
$\zeta$ as $I$ times the complex conjugate of the phase of the $\alpha$
symplectic component.)  But the effect of this multiplication is to just
induce a gauge transformation on the vector potential $\vec A(\vec r \,)$,
and a corresponding quaternion automorphism transformation of the scalar
potential $\tilde U$.  Hence we can always pick a gauge for the
potentials  in the shell model Hamiltonian which makes the {\it ground
state} wave function (but not simultaneously the wave functions of
higher excited states) quaternion imaginary.  Working in this
gauge we get the first part of Rule 1, with the identification
$$a_{0}(\vec r \,) =- \vec a(\vec r \,) \cdot \vec E \> \eqno(10b) $$
for the wave function $\vec a(\vec r \,)$ introduced above.
Although it may seem objectionable to have to assume a specific gauge to
formulate the model, this feature was also present in the original form of
the BCS theory of superconductivity, and this analogy suggests that
as in the
case of superconductivity, gauge invariance should be restored
by the proper inclusion of collective effects.

Finally, we note that since the three quasiparticle approximation
to the three  body ground state
wave function is not exact, and since the shell model itself
represents an approximation,  there are residual forces which act on
the composites.  These residual forces can, in principle, give rise to
 the gauge
fields of the standard model which act on the quarks and leptons.  We
have argued in Ref. [7] that because the left algebra structure of
quaternionic Hilbert space can give rise to multi-quaternion algebras,
it is possible to build up larger effective gauge groups than the
underlying $SU(2) \times SU(2)$ preonic gauge group.  However, since the
underlying gauge group is vector-like, it seems reasonable to suppose
that any larger gauge groups kinematically generated from it will
still not couple
spin $\uparrow$ to $\downarrow$ components, which is the basis for the
identification of spin-associated tripling with family structure in
Rule 3.  In  this picture the
chiral structure of the weak interactions is not fundamental, but must
arise from spontaneous symmetry breaking at a scale lying at or below
the preonic energy scale.

Although the above sketch of how one might attempt to justify
Rules 1-3 in a more
fundamental theory is only schematic, and leaves much to be
done, I believe that it is consistent with both our current
 experimental knowledge and with
the known properties of complex and quaternionic quantum mechanics.

I developed the ideas reported here in the course of giving a series
of lectures covering material in Ref. [7] to a group of Institute for
Advanced Study members, Y.M. Cho, C. Moreira, R. Narayanan, and Y.-J. Ng,
and Princeton University graduate students, A. Millard and J. Weckel.
Their criticism, corrections of my errors, and healthy skepticism of
features of earlier versions were essential in my arriving at
the formulation presented here, and I owe them my profound thanks.  I
wish to thank W. Bardeen and E. Witten for conversations about
electroweak phenomenology and the couplings of spin 3/2 particles.  I
also wish to acknowledge the support of the Department of Energy, under
Grant\# DE-FG02-90ER40542.
\vfill
\eject
\leftline {\bf References}
\smallskip
\noindent
[1] H. Harari, Phys. Lett. {\bf 86b}, 83 (1979).
\smallskip
\noindent
[2] M.A. Shupe, Phys. Lett. {\bf 86b}, 87 (1979).
\smallskip
\noindent
[3] S.L. Adler, Phys. Rev. {\bf D21}, 2903 (1980).
\smallskip
\noindent
[4] S.L. Adler, {\it Family replication in the Harari-Shupe composite
model}, IASSNS-HEP-87/33, unpublished (1987).
\smallskip
\noindent
[5] L.P. Horwitz and L.C. Biedenharn, Ann. Phys. {\bf157}, 432 (1984).
\smallskip
\noindent
[6] S.L. Adler, Phys. Rev. {\bf D37}, 3654 (1988).
\smallskip
\noindent
[7] S.L. Adler, {\it Quaternionic Quantum Mechanics and Quantum Fields},
Oxford University Press, in press.
\smallskip
\noindent
[8] A. Govorkov, Teoret. i Mat. Fiz. {\bf 68}, 381 (1986); English
translation in Theor. Math. Phys. {\bf 68}, 893 (1987); A.B. Govorkov,
Teoret. i Mat. Fiz. {\bf 69}, 69 (1986); English translation in Theor.
Math. Phys. {\bf 69}, 1007 (1987).
\smallskip
\noindent
[9] D. Faiman and A. Hendry, Phys. Rev. {\bf 173}, 1720 (1968).
\smallskip
\noindent
[10] R.P. Feynman, M. Kislinger, and F. Ravndal, Phys. Rev. {\bf D11},
2706 (1971), Sec. 1 of the Appendix.
\smallskip
\noindent
[11] G. Altarelli, unpublished talk given at the conference {\it Around
the Dyson Sphere}, Princeton, NJ, April 8-9, 1994.
\smallskip
\noindent
[12] S.L. Adler, Nucl. Phys. {\bf B145}, 195 (1994), Sec. 4.
\smallskip
\noindent
[13] S.L. Adler, {\it Scattering Theory in Quaternionic Quantum
Mechanics}, in A. Das., ed., {\it From Symmetries to Strings: Forty
Years of Rochester Conferences.}  World Scientific, Singapore, 1990.
\vfill
\eject
\leftline{Table 1}
\smallskip
\noindent
The four basic quaternionic quasiparticle annihilation operators used to
construct the annihilation operators for three quasiparticle composite
states centered at $\vec R$.  The annihilation operators $t_{\uparrow
,\downarrow},\,v_{\uparrow ,\downarrow}$ are conventional fermion
annihilation operators, and are real numbers with respect to the
quaternion algebra
$E_AE_B=-\delta_{AB}+\sum_C~\varepsilon_{ABC}E_C,~A,B,C=1,2,3$.  The
spin and flavor independent coefficient $\vec a(\vec r \,)\cdot\vec
E=\sum_A~a_A(\vec r \,)E_A$ is a quaternion imaginary internal ground state
wave function.  The notation of the abbreviated forms is used in Tables
2 and 3.
\bigskip
\leftline{Full form:}
$$
\eqalign{T_\uparrow(\vec R)
&=\int~d^3r\,\vec a(\vec r\,)\cdot\vec E\>t_\uparrow(\vec r+\vec R)\cr
T_\downarrow(\vec R)
&=\int~d^3r\,\vec a(\vec r\,)\cdot\vec E\>t_\downarrow(\vec r+\vec R)\cr
V_\uparrow(\vec R)
&=\int~d^3r\,\vec a(\vec r\,)\cdot\vec E\>v_\uparrow(\vec r+\vec R)\cr
V_\downarrow(\vec R)
&=\int~d^3r\,\vec a(\vec r\,)\cdot\vec E\>v_\downarrow(\vec r+\vec R)\cr}
$$
\bigskip
\leftline{Abbreviated form:}
$$
\eqalign{T_{\uparrow ,\downarrow}
&=\sum_1~\vec a(1)\cdot\vec E\>t_{\uparrow ,\downarrow}(1)\cr
V_{\uparrow ,\downarrow}
&=\sum_1~\vec a(1)\cdot\vec E\>v_{\uparrow ,\downarrow}(1)\cr}
$$
\vfill
\eject
\leftline{Table 2}
\smallskip
\noindent
Three quasiparticle composites with a totally antisymmetric internal
wave function structure, classified by spin, charge, and interaction
type.  For each charge 1 and charge 2/3 state in the table, there is a
corresponding charge 0 and charge 1/3 state obtained by the interchange
$T \leftrightarrow V, \> t \leftrightarrow v$.  Similarly, negative helicity
states are obtained
from positive helicity ones by the interchange $\uparrow \leftrightarrow
\downarrow$.
\bigskip
\item{(1)}  Spin 3/2, charge 1, lepton
\medskip
\item\item{helicity 3/2}
$$
T_\uparrow T_\uparrow T_\uparrow = -\sum_{1,2,3}~\vec a(1)\times\vec
a(2)\cdot \vec a(3)\>t_\uparrow(1)t_\uparrow(2)t_\uparrow(3)
$$
\smallskip
\item\item{helicity 1/2}
$$
T_\uparrow T_\downarrow T_\uparrow = -\sum_{1,2,3}~\vec
a(1)\times \vec a(2)\cdot \vec
a(3)\>t_\uparrow(1)t_\downarrow(2)t_\uparrow(3)
$$
\medskip
\item{(2)}  Spin 3/2, charge 2/3, lepton
\medskip
\item\item{helicity 3/2}
$$
T_\uparrow V_\uparrow T_\uparrow =-\sum_{1,2,3}~\vec a(1)\times
\vec a(2)\cdot \vec a(3)\> t_\uparrow(1)v_\uparrow(2)t_\uparrow (3)
$$
\smallskip
\item\item{helicity 1/2}
$$
\eqalign{
&\qquad\qquad\qquad T_\uparrow V_\uparrow T_\downarrow
+ T_\uparrow V_\downarrow T_\uparrow + T_\downarrow V_\uparrow
T_\uparrow\cr
&=-\sum_{1,2,3}~\vec a(1)\times \vec a(2)\cdot \vec a(3)\>[t_\uparrow
(1)v_\uparrow (2)t_\downarrow (3)+
t_\uparrow(1)v_\downarrow(2)t_\uparrow(3)+
t_\downarrow(1)v_\uparrow(2)t_\uparrow(3)]\cr}
$$
\medskip
\item{(3)} Spin 1/2, charge 2/3, lepton
\medskip
\item\item{helicity 1/2}
$$
T_\uparrow V_\uparrow T_\downarrow +T_\downarrow V_\uparrow T_\uparrow
-2T_\uparrow V_\downarrow T_\uparrow = -2\sum_{1,2,3}~\vec a(1)\times
\vec a(2)\cdot \vec a(3)\>[t_\uparrow(1)
v_\uparrow(2)t_\downarrow(3)-t_\uparrow(1)v_\downarrow(2)t_\uparrow(3)]
$$
\vfill
\eject
\overfullrule=0pt

\leftline{Table 3}
\smallskip
\noindent
Three quasiparticle composites with a mixed symmetry internal wave
function, classified by spin, charge, and interaction type.  The index
$A$ takes the values $1,2,3$, resulting in three copies of each state.
This tripling is interpreted as corresponding to three colors when it
arises from $t,v$ reorderings, and as corresponding to three families
when it arises from spin $\uparrow,\downarrow$ reorderings.  For each
charge 1 and charge 2/3 state in the table, there is a corresponding
charge 0 and charge 1/3 state obtained by the interchange
$T\leftrightarrow V, \> t \leftrightarrow v$.  Similarly, negative helicity
states are obtained
from positive helicity ones by the interchange $\uparrow \leftrightarrow
\downarrow$
\medskip
\item{(1)}  Spin 3/2, charge 2/3, quark
\medskip
\item\item{helicity 3/2}
$$
T_\uparrow T_\uparrow V_\uparrow - V_\uparrow T_\uparrow T_\uparrow =
2\sum_{1,2,3,A}~[\vec a(1)\cdot \vec a(3) a_A(2)-\vec a(2)\cdot\vec
a(3)a_A(1)]\,t_\uparrow(1)t_\uparrow(2)v_\uparrow(3) \, E_A
$$
\medskip
\item\item{helicity 1/2}
$$
\eqalign{
&\qquad T_\downarrow T_\uparrow V_\uparrow + T_\uparrow T_\downarrow
V_\uparrow + T_\uparrow T_\uparrow V_\downarrow - (V_\uparrow
T_\downarrow T_\uparrow+V_\uparrow T_\uparrow T_\downarrow +
V_\downarrow T_\uparrow T_\uparrow)\cr
&=2\sum_{1,2,3,A}~[\vec a(1)\cdot \vec a(3)a_A(2)-\vec a(2)\cdot\vec
a(3)a_A(1)][t_\downarrow(1)t_\uparrow(2)v_\uparrow(3)+t_\uparrow(1)
t_\downarrow(2)v_\uparrow(3)+t_\uparrow(1)t_\uparrow(2) \break
v_\downarrow(3)] \, E_A \cr}
$$
\medskip
\item{(2)}  Spin 1/2, charge 1, lepton
\medskip
\item\item{helicity 1/2}
$$
\eqalign{
&TTT \times (\uparrow \uparrow \downarrow \break
-\downarrow \uparrow \uparrow) = T_\uparrow T_\uparrow \break
T_\downarrow -T_\downarrow T_\uparrow  T_\uparrow \cr
=& 2\sum_{1,2,3,A}~[\vec a(1)\cdot \vec a(3)a_A(2)-\vec a(2)\cdot \vec a(3)
a_A(1)]\,t_\uparrow(1) t_\uparrow(2) t_\downarrow(3) \, E_A \cr
}
$$
\medskip
\item{(3)} Spin 1/2, charge 2/3, quark
\medskip
\item\item{helicity 1/2}
$$
\eqalign{
&(TTV+VTT) \times (\uparrow \uparrow \downarrow - \break
\downarrow \uparrow \uparrow) \cr
=& T_\uparrow T_\uparrow
V_\downarrow+ V_\uparrow T_\uparrow T_\downarrow \break
- T_\downarrow T_\uparrow V_\uparrow - V_\downarrow \break
T_\uparrow T_\uparrow\cr
=&2\sum_{1,2,3,A}~\{-\vec a(1)\cdot \vec a(2)a_A(3) \, \break
t_\uparrow(1) t_\downarrow(2) v_\uparrow(3) \cr
&\qquad +[\vec a(1) \cdot \vec a(3) a_A(2)-\vec a(2) \cdot \break
\vec a(3) a_A(1)] [t_\uparrow(1) t_\uparrow(2) v_\downarrow(3) \break
-t_\uparrow(1) t_\downarrow(2) v_\uparrow(3)] \} \, E_A \cr
}
$$
\medskip
\item\item{Spin 1/2, charge 2/3, quark}
\medskip
\item\item{helicity 1/2}
$$\eqalign{
&TVT \times (\uparrow \uparrow \downarrow -\downarrow \uparrow \break
\uparrow) \cr
=&T_\uparrow V_\uparrow T_\downarrow-T_\downarrow V_\uparrow \break
T_\uparrow \cr
=&2\sum_{1,2,3,A}~[\vec a(2)\cdot \vec a(3)a_A(1) \break
+\vec a(1) \cdot \vec a(3) a_A(2) - \vec a(1) \cdot \vec a(2) \break
a_A(3)] \, t_\uparrow(1)t_\downarrow(2)v_\uparrow(3) \, E_A \cr
}
$$
\medskip
\item\item{Spin 1/2, charge 2/3, quark}
\medskip
\item\item{helicity 1/2}
$$
\eqalign{
&(TTV-VTT) \times (\uparrow \uparrow \downarrow + \downarrow \break
\uparrow \uparrow -2 \uparrow \downarrow \uparrow) \cr
=&  T_\uparrow T_\uparrow V_\downarrow - V_\uparrow T_\uparrow \break
T_\downarrow + T_\downarrow T_\uparrow V_\uparrow - V_\downarrow \break
T_\uparrow T_\uparrow -2 T_\uparrow T_\downarrow V_\uparrow \break
+2 V_\uparrow T_\downarrow T_\uparrow \cr
=&2\sum_{1,2,3,A}~\{ 3 \vec a(1)\cdot\vec a(2)a_A(3) \, \break
t_\uparrow(1) t_\downarrow(2) v_\uparrow(3)  \cr  \break
&\qquad +[\vec a(1) \cdot \vec a(3) a_A(2)-\vec a(2) \cdot \break
\vec a(3) a_A(1)] [t_\uparrow(1) t_\uparrow(2) v_\downarrow(3) \break
-t_\uparrow(1) t_\downarrow(2) v_\uparrow(3)] \} \, E_A \cr
}
$$

\bye